\documentclass[sigconf]{acmart}

\renewcommand\footnotetextcopyrightpermission[1]{} 
\newcommand\blfootnote[1]{%
	\begingroup
	\renewcommand\thefootnote{}\footnote{#1}%
	\addtocounter{footnote}{-1}%
	\endgroup
}

\usepackage{graphicx}

\usepackage{amsmath}
\usepackage{booktabs}

\usepackage{suffix}
\usepackage{color}
\usepackage{dsfont}
\usepackage{xspace}
\usepackage{amsfonts}
\usepackage{amsmath}
\usepackage{algorithm}
\usepackage[shortlabels]{enumitem}
\usepackage[noend]{algpseudocode}
\usepackage{flushend}
\usepackage{tabularx}

\newcommand{\squishlist}{\begin{list}{$\bullet$}
  { \setlength{\itemsep}{0pt}
     \setlength{\parsep}{3pt}
     \setlength{\topsep}{3pt}
     \setlength{\partopsep}{0pt}
     \setlength{\leftmargin}{1.5em}
     \setlength{\labelwidth}{1em}
     \setlength{\labelsep}{0.5em} } }
\newcommand{\squishend}{
\end{list}  }

\newcommand{\edit}[1]{#1}
\newcommand{\rcomment}[1]{\textcolor{red}{\textbf{[#1]}}}

\newcommand{\neighbors}{\ensuremath{\Delta}\xspace}

\newcommand{\natalie}{{\sc Natalie}\xspace}
\newcommand{\netalign}{{\sc Net\-Align\-MP}{\tt ++}\xspace}
\newcommand{\isorank}{{\sc IsoRank}\xspace}
\newcommand{\isorankn}{{\sc IsoRankN}\xspace}
\newcommand{\graal}{{\sc Graal}\xspace}
\newcommand{\lgraal}{{\sc L-Graal}\xspace}

\newcommand{\similarity}{\ensuremath{\sigma}\xspace}

\newcommand{\p}[1]{\ensuremath{#1}\%}
\newcommand{\tm}{{\sc Top\-Matchings}\xspace}
\newcommand{\gm}{{\sc Gibbs\-Matchings}\xspace}
\newcommand{\margin}{{\sc Margin}\xspace}
\newcommand{\lccl}{{\sc lccl}\xspace}

\newcommand{\betw}{{\sc Bet\-ween\-ness}\xspace}
\newcommand{\random}{{\sc Random}\xspace}

\newcommand{\oracleq}{\ensuremath{\text{\sc Oracle\-Query}}\xspace}
\newcommand{\Cands}{\ensuremath{\mathcal{C}}\xspace}
\newcommand{\bm}[1]{{{\sc BipartiteMatching}(#1)}\xspace}
\newcommand{\alignment}{{\sc Net\-work\-Align\-ment}$\left(G_s, G_t \mid 
\Cands\right)$\xspace}

\newcommand{\certainty}{\ensuremath{\mathrm{Cert}}}

\newcommand{\calM}{\ensuremath{\mathcal{M}}}

\newcommand{\argmin}{\mathrm{argmin}}

\newcommand{\spara}[1]{\smallskip\noindent{\bf{#1}}}

\newcommand{\pr}[1]{\left(#1\right)}
\newcommand{\fpr}[1]{\mathopen{}\left(#1\right)}

\newcommand{\np}{\textbf{NP}}

\DeclareRobustCommand{\dispfunc}[2]{%
  \ensuremath{%
  \ifthenelse{\equal{#2}{}}%
    {{#1}}%
    {{#1}\fpr{#2}}}}

\newcommand{\dd}{\ensuremath{K}}

\newcommand{\dist}[1][(p, q)]{\def\ArgI{{#1}}\distRelay}
\newcommand{\distRelay}[2][]{\dispfunc{\dd^{\ArgI}_{#1}}{#2}}
\WithSuffix\newcommand\dist*[2][]{\dist[][#1]{#2}}

\newcommand{\prob}[1]{P\pr{#1}}

\newtheorem{problem}{Problem}

\usepackage{tikz,pgfplots,pgfplotstable}
\usetikzlibrary{positioning,fit,shapes}
\usetikzlibrary{decorations.pathreplacing}

\pgfdeclarelayer{background}
\pgfdeclarelayer{foreground}
\pgfsetlayers{background,main,foreground}

\makeatletter
\tikzset{multicircle/.style  args={#1, #2}{%
 alias=tmp@name, %
  postaction={%
    insert path={
     \pgfextra{%
     \pgfpointdiff{\pgfpointanchor{\pgf@node@name}{center}}%
                  {\pgfpointanchor{\pgf@node@name}{east}}%
     \pgfmathsetmacro\insiderad{\pgf@x}%
        \fill[white] (\pgf@node@name.center)  circle (\insiderad-\pgflinewidth);%
        \draw[#2] (\pgf@node@name.center)  circle (\insiderad-\pgflinewidth);%
        \fill[#2] (\pgf@node@name.center)  -- ++(0:\insiderad-\pgflinewidth) arc (0:#1:\insiderad-\pgflinewidth)--cycle;%
        }}}}}
\makeatother

\definecolor{yafaxiscolor}{rgb}{0.3, 0.3, 0.3}
\definecolor{yafcolor1}{rgb}{0.4, 0.165, 0.553}
\definecolor{yafcolor2}{rgb}{0.949, 0.482, 0.216}
\definecolor{yafcolor3}{rgb}{0.47, 0.549, 0.306}
\definecolor{yafcolor4}{rgb}{0.925, 0.165, 0.224}
\definecolor{yafcolor5}{rgb}{0.141, 0.345, 0.643}
\definecolor{yafcolor6}{rgb}{0.965, 0.933, 0.267}
\definecolor{yafcolor7}{rgb}{0.627, 0.118, 0.165}
\definecolor{yafcolor8}{rgb}{0.878, 0.475, 0.686}
\definecolor{yafcolor9}{rgb}{0.965, 0.733, 0.767}

\newlength{\yafaxispad}
\newlength{\yaftlpad}
\newlength{\yaflabelpad}
\newlength{\yafaxiswidth}
\newlength{\yafticklen}
\makeatletter
\def\pgfplots@drawtickgridlines@INSTALLCLIP@onorientedsurf#1{}
\makeatother

\newcommand{\yafdrawxaxis}[2]{
	\pgfplotstransformcoordinatex{#1}\let\xmincoord=\pgfmathresult 
	\pgfplotstransformcoordinatex{#2}\let\xmaxcoord=\pgfmathresult 
	\pgfsetlinewidth{\yafaxiswidth} 
	\pgfsetcolor{yafaxiscolor}
	\pgfpathmoveto{\pgfpointadd{\pgfpointadd{\pgfplotspointrelaxisxy{0}{0}}{\pgfqpointxy{\xmincoord}{0}}}{\pgfqpoint{-0.5\yafaxiswidth}{\yafaxispad}}}
	\pgfpathlineto{\pgfpointadd{\pgfpointadd{\pgfplotspointrelaxisxy{0}{0}}{\pgfqpointxy{\xmaxcoord}{0}}}{\pgfqpoint{0.5\yafaxiswidth}{\yafaxispad}}}
	\pgfusepath{stroke}

}
\newcommand{\yafdrawyaxis}[2]{
	\pgfplotstransformcoordinatey{#1}\let\ymincoord=\pgfmathresult 
	\pgfplotstransformcoordinatey{#2}\let\ymaxcoord=\pgfmathresult 
	\pgfsetlinewidth{\yafaxiswidth} 
	\pgfsetcolor{yafaxiscolor}
	\pgfpathmoveto{\pgfpointadd{\pgfpointadd{\pgfplotspointrelaxisxy{0}{0}}{\pgfqpointxy{0}{\ymincoord}}}{\pgfqpoint{\yafaxispad}{-0.5\yafaxiswidth}}}
	\pgfpathlineto{\pgfpointadd{\pgfpointadd{\pgfplotspointrelaxisxy{0}{0}}{\pgfqpointxy{0}{\ymaxcoord}}}{\pgfqpoint{\yafaxispad}{0.5\yafaxiswidth}}}
	\pgfusepath{stroke}
}

\pgfplotscreateplotcyclelist{yaf}{%
{yafcolor1,mark options={scale=0.75},mark=o}, 
{yafcolor2,mark options={scale=0.75},mark=square},
{yafcolor3,mark options={scale=0.75},mark=triangle},
{yafcolor4,mark options={scale=0.75},mark=o},
{yafcolor5,mark options={scale=0.75},mark=o},
{yafcolor6,mark options={scale=0.75},mark=o},
{yafcolor7,mark options={scale=0.75},mark=o},
{yafcolor8,mark options={scale=0.75},mark=o}} 

\pgfplotsset{axis y line=left, axis x line=bottom,
	tick align=outside,
	compat = 1.3,
	tickwidth=\yafticklen,
	clip = false,
	every axis title shift = 0pt,
    x axis line style= {-, line width = 0pt, opacity = 0},
    y axis line style= {-, line width = 0pt, opacity = 0},
    x tick style= {line width = \yafaxiswidth, color=yafaxiscolor, yshift = \yafaxispad},
    y tick style= {line width = \yafaxiswidth, color=yafaxiscolor, xshift = \yafaxispad},
    x tick label style = {font=\scriptsize, yshift = \yaftlpad},
    y tick label style = {font=\scriptsize, xshift = \yaftlpad},
    every axis y label/.style = {at = {(ticklabel cs:0.5)}, rotate=90, anchor=center, font=\scriptsize, yshift = -\yaflabelpad},
    every axis x label/.style = {at = {(ticklabel cs:0.5)}, anchor=center, font=\scriptsize, yshift = \yaflabelpad},
    x tick label style = {font=\scriptsize, yshift = 1pt},
    grid = major,
    major grid style  = {dash pattern = on 1pt off 3 pt},
	every axis plot post/.append style= {line width=\yafaxiswidth} ,
	legend cell align = left,
	legend style = {inner sep = 1pt, cells = {font=\scriptsize}},
	legend image code/.code={%
		\draw[mark repeat=2,mark phase=2,#1] 
		plot coordinates { (0cm,0cm) (0.15cm,0cm) (0.3cm,0cm) };%
	} 
}

\begin{document}
\title{Active Network Alignment: A Matching-Based Approach}

\author{Eric Malmi}
\affiliation{%
  \institution{Aalto University}
  \city{Espoo}
  \country{Finland}
}
\email{eric.malmi@aalto.fi}

\author{Aristides Gionis}
\affiliation{%
	\institution{Aalto University}
	\city{Espoo}
	\country{Finland}
}
\email{aristides.gionis@aalto.fi}

\author{Evimaria Terzi}
\affiliation{%
  \institution{Boston University}
  \city{Boston}
  \country{USA}
}
\email{evimaria@cs.bu.edu}


\begin{abstract}
Network alignment is the problem of matching the nodes of two graphs, 
maximizing the similarity of the matched nodes and the edges between them. 
This problem is encountered in a wide array of applications---from biological 
networks to social networks to ontologies---where multiple networked data 
sources need to be integrated. 
Due to the difficulty of the task, 
an accurate alignment can rarely be found without human assistance.
Thus, it is of great practical importance to develop network alignment 
algorithms that can optimally leverage experts who are able to provide 
the correct alignment for a small number of nodes. Yet, only a handful of 
existing works address this active network alignment setting.

The majority of the existing active methods focus on \textit{absolute 
queries} (``are nodes $a$ and $b$ the same or not?''), 
whereas we argue that it 
is generally easier for a human expert to answer \textit{relative queries} 
(``which node in the set $\{b_1, \ldots, b_n\}$ is the most similar to node $a$?''). 
This paper introduces two novel relative-query strategies, \tm and \gm, 
which can be applied on top of any network alignment method that 
constructs and solves a bipartite matching problem. 
Our methods identify the most informative nodes to query by sampling the
matchings of the bipartite graph
associated to the network-alignment instance.

We compare the proposed approaches to several commonly-used query 
strategies and perform experiments on both synthetic and real-world datasets. 
Our sampling-based strategies yield the highest overall performance, 
outperforming all the baseline methods by more than 15 percentage points in 
some cases. 
In terms of accuracy, \tm and \gm perform comparably. However, \gm is significantly more scalable, but it also requires hyperparameter tuning for a temperature parameter.
\blfootnote{This is a pre-print of an article appearing at CIKM 2017.}
\end{abstract}


\maketitle

\section{Introduction} 
\label{sec:intro}

The network-alignment problem, also known as graph matching \cite{yan16} or graph reconciliation \cite{korula2014}, asks to find a 
matching between the nodes of two graphs so that both 
$(i)$ node-attribute similarities and $(ii)$ structural similarities between the matched nodes are maximized.
This is an ubiquitous problem with application areas ranging from biological networks \cite{clark2014} to social networks \cite{goga2015reliability,zhang2015multiple}, ontologies \cite{shvaiko2013}, and image matching in computer vision \cite{ConteFSV04}. For instance, in the case of social networks, one might be interested in aligning the friendship graphs of two social-networking services in order to suggest new friends for the users.

Typically, some of the network nodes are easy to align auto\-matically if they, 
for example, share a unique name. 
Other nodes can be considerably more ambiguous 
and thus fully-automatic methods are likely to align them incorrectly.
In the \emph{active} version of the network-alignment problem, 
such difficult cases are redirected to human experts who act as oracles.
In this way, the alignment process is judiciously enhanced by involving  
humans in the loop.

The idea of algorithms 
that select which data to be labeled in order to improve accuracy is not new;  
in fact, this is the main focus of \emph{active learning}.  
Research in this area aims to identify
effective ways for utilizing access to labeling oracles, such as human experts. 
Although there is a lot of research in active learning for classification or clustering problems, 
few studies have tackled the problem of \emph{active network alignment}.

To the best of our knowledge,
active network-alignment methods
appear mostly in the domain of ontology matching~\cite{jimenez2012,sarasua2012,shi2009}.
These methods ask the human experts to assess
whether two given nodes are a match or not,
and thus, focus on identifying the most useful pair of nodes to query.
The limitation of this approach is that absolute yes/no questions can be 
very hard to answer for a human if no context about alternative candidate matches is provided.

In this paper, we obtain human feedback, where
questions are asked
in the following form:
\emph{``Given node $v$ and a set of candidate matches $C$, which node in 
$C$ is the most likely match for $v$?''}. 
To answer such \emph{relative} questions, 
an expert needs to make only comparative judgments, 
which are less challenging for humans \cite{laming2003human},
despite the fact that the expert needs to consider more nodes at once.
Additionally, the expert may be given an opportunity to say that none of the 
candidate matches is correct. In this scenario, the querying is more similar to 
the absolute querying scheme, 
but it may still be easier for the 
expert since more context is provided to answer the 
question.\footnote{An 
interesting parallel to the absolute vs.\ relative querying issue is found in 
the psychology literature regarding eyewitness identifications.
Some experimental studies show that simultaneous lineups, where suspects are 
shown to an eyewitness simultaneously, result in a higher true positive rate, 
whereas sequential lineups result in a lower false positive 
rate~\cite{carlson2008}.} 
Although in certain scenarios the absolute querying scheme may be more appropriate, this work focuses only on comparing different relative approaches.

Given the above relative querying scheme, 
our framework for active network alignment is based on a novel algorithmic idea 
for identifying the best questions to ask to the experts. 
Since access to experts is typically costly, 
the objective is to maximize the alignment accuracy for a given
number of queries.

A well-established approach in active learning is to label the data points
for which the current model 
is \emph{least certain} as to what the correct output should be \cite{settles2010}. Accordingly, we introduce novel ways of quantifying the 
\emph{uncertainty} introduced by each node
in the network-alignment process.
Using these measures we ask queries that resolve most of the uncertainty,
and therefore,
only few queries suffice to obtain an alignment of high accuracy.

In the classic (non-active) network-alignment problem
the input consists of two networks: 
a \emph{source} network $G_s=(V_s,E_s)$
and a \emph{target} network $G_t=(V_t,E_t)$.
Finding an alignment is often reduced into the problem of finding a 
\emph{matching} on a weighted bipartite graph $H=(V_s,V_t,E_h)$, where the 
weights incorporate both attribute and structural similarities. Examples of 
such methods include 
{\lgraal}~\cite{malod2015graal},
{\natalie}~\cite{el2015,klau2009}, 
{\netalign}~\cite{bayati2013}, and
{\isorank}~\cite{singh2008}.

In this paper we propose a new approach for active network alignment, 
which can be employed on top 
of any matching-based (non-active) network-alignment method.
The main idea is to sample a set of matchings $\calM_\ell$, 
use the resulting distribution to quantify the certainty of each node, 
and identify which node to query based on this certainty. 
We study and experiment with two alternative methods for 
obtaining a set of sampled matchings, 
\gm, where Gibbs sampling is employed to sample matchings according to their score, and 
\tm, where the top-$\ell$ matchings are taken in the sample.
We also experiment with different methods for estimating node certainty, 
using entropy, the expected certainty of the unlabeled nodes, 
as well as a consensus-based criterion.

Our experiments with real and synthetic data show that the proposed strategies perform
consistently well and in some cases they outperform our baseline methods 
by more than 15\% in accuracy. 
The baseline methods include three previously
proposed query strategies as well as random querying.
When comparing the two proposed methods, 
\gm and \tm, we obtain a robustness-scalability trade-off:
\gm is significantly more scalable but it is also sensitive to the choice of a temperature parameter $\beta$.

Our main contributions are summarized as follows.

\squishlist
\item We formalize a relative-judgment framework for active network alignment.

\item We develop an active-querying framework, which can be employed on top of any network-alignment
method that finds an alignment using maximum-weight bipartite matching. 
Indeed, several state-of-the-art network-alignment methods follow this approach. 
We also explore two algorithms that instantiate this framework: {\gm} and {\tm}.

\item We conduct experiments with real and synthetic datasets, demonstrating 
the superiority of our algorithms compared to several previously proposed baseline methods.
We also show that our algorithms can be parallelized without significantly compromising the 
accuracy of the methods.
\item The code and the data used in the experiments are publicly available 
at: \url{https://github.com/ekQ/active-network-alignment}
\squishend  

\spara{Roadmap:} The paper is organized as follows.
In Section~\ref{section:relatedwork} we review the related work.
Our problem formulation is provided in 
Section~\ref{section:problem-formulation} and our algorithm
in Section~\ref{section:algorithms}. 
Section~\ref{section:experiments}
presents the experimental evaluation of our method and a comparison with
baselines. 
We conclude in Section~\ref{section:conclusions}.

\section{Related work}
\label{section:relatedwork}

Numerous methods have been developed for the non-active network-alignment problem. 
The problem has drawn particular attention in the bioinformatics domain 
\cite{el2015,hashemifar2014hubalign,klau2009,kuchaiev2010topological,liao2009isorankn,malod2015graal,singh2008}, 
due to the interest in the the task of matching protein-protein interaction networks; 
a recent survey in the area is provided by Elmsallati et al.~\cite{elmsallati2016global}.
Non-active network alignment methods are classified according to how the matching cost is defined 
and how they algorithmically proceed to finding a solution.
For example, {\isorank}~\cite{singh2008} and {\isorankn}~\cite{liao2009isorankn}, 
which are among the earliest-developed methods, 
utilize a PageRank-type computation to recursively compute node similarity 
via the similarity of the nodes' neighbors.
\natalie~\cite{el2015,klau2009}
formulates the alignment task as a quadratic assignment problem, 
which it then solves using Lagrangian relaxation 
combined with a subgradient optimization; 
more details are given in Section~\ref{section:natalie-netalign}. 
The {\graal} \cite{kuchaiev2010topological,kuchaiev2011integrative,malod2015graal} 
family of alignment methods 
enhance the matching scoring function with topological similarity features, 
such as graphlet degree signatures~\cite{prvzulj2007biological}.
{\lgraal}~\cite{malod2015graal} is a recent algorithm in the {\graal} family, 
which incorporates the Lagrangian-relaxation framework of \natalie, 
and is shown to outperform several other state-of-the-art methods.

In addition to bioinformatics, the non-active network-alignment problem 
has been studied in different application areas, 
such as ontology matching \cite{aumueller2005, doan2004}
and social-network matching \cite{korula2014}.

As discussed earlier, 
our active network-alignment framework
can be employed on top of any non-active method 
that maps the alignment problem into a weighted bipartite graph-matching problem.
Many of the methods discussed above fall in this category, 
e.g., 
{\lgraal}~\cite{malod2015graal},
{\natalie}~\cite{el2015,klau2009}, 
{\netalign}~\cite{bayati2013}, and
{\isorank}~\cite{singh2008}.
In our experimental evaluation we use {\natalie} and {\netalign}, 
which are state-of-the-art methods that have been shown to have a robust performance 
in several independent studies~\cite{bayati2013, clark2014,el2015,malod2015graal}.

The problem of active network alignment has been previously studied by 
Cort\'{e}s and Serratosa~\cite{cortes2013}. Compared to our work, they focus on 
a more limited class of network-alignment methods that return a probability 
matrix for different matches, making the quantification of uncertainty more 
straightforward. However, some of their query strategies are also applicable to 
our setting and thus, in our experiments, we adopt two baseline strategies from their work, namely 
\lccl and \margin.
Another closely-related line of work is active ontology matching. 
Shvaiko and Euzenat \cite{shvaiko2013} list it as one of the important areas 
for future work in their recent survey on ontology matching. 
Existing work on active ontology matching focuses on absolute queries~\cite{jimenez2012,sarasua2012,shi2009},
and thus, not directly comparable with our approach.

Active-learning approaches have been developed also for other related problems. 
Charlin et al.~\cite{charlin2012} develop an active-learning method for 
many-to-many matching problems encountered in recommender systems, comparing 
different absolute-querying strategies. Another area where active learning has 
been studied and where data is in a network format is the inference 
problem for Gaussian fields \cite{macskassy2009,zhu2003}. Macskassy 
\cite{macskassy2009} finds an empirical risk minimization ({\sc erm}) to be the best 
method to find the next instance to query but due to the high computational 
complexity of {\sc erm}, he proposes to use the betweenness centrality as a filter to 
select a subset of nodes for which {\sc erm} is applied to. 
Finally, Bilgic et al.~\cite{bilgic2010} introduce an active learning method for collectively 
classifying networked data.

\section{Problem~formulation}
\label{section:problem-formulation}

Before defining the 
{\em active network alignment} problem, 
we first discuss the non-active version of the problem.

\spara{Network alignment:}
In the standard network-alignment problem 
we consider two input graphs $G_s=(V_s, E_s)$ and $G_t = (V_t, E_t)$, 
the \emph{source} and the \emph{target} graph. 
The adjacency matrices of the two graphs 
are denoted by $A_s$ and $A_t$, respectively.
Throughout we assume that the smaller graph is aligned to the larger one, that 
is, $|V_s|\leq |V_t|$.
In addition, we consider that a similarity function 
$\similarity:V_s\times V_t\rightarrow \mathbb{R}$ 
is available, 
which measures the similarity between pairs of nodes $i\in V_s$ and $j\in V_t$.
The similarity function $\similarity$ depends on the application at hand. 

The objective of the network-alignment problem 
is to find a {\em matching} between the nodes of the two networks. 
More specifically, we want to align the 
nodes of the source graph $V_s$ to the nodes of the target graph~$V_t$.
Formally, we want to find 
$M=\{(i,j)\}\subseteq V_s\times V_t$
so that each node of $V_s$ and $V_t$ appears at most one time in~$M$.
A high-quality alignment should satisfy the following properties: 
($i$)~nodes in $V_s$ should match to similar nodes in $V_t$, according to 
$\similarity$,
and ($ii$)~the endpoints of each edge in $E_s$ should match to nodes in $V_t$ 
that are connected by edges in $E_t$. 

For each node $v\in V_s$
we consider a set of {\em candidate matching nodes } $\Cands_v\subseteq V_t$.
Those are the only nodes in $V_t$ that $v$ can be matched to.
In this paper we assume that the sets $\Cands_v$ are given.
In practice, the sets $\Cands_v$ are computed by a simple heuristic, 
e.g., considering all nodes in $V_t$ whose feature-based similarity to 
$v\in V_s$ exceeds a certain threshold.
When the candidate sets are small compared to $V_t$, the problem is sometimes 
called {\em sparse network alignment}.

{\em Global} network-alignment methods
formulate an objective function that captures the above properties, 
and then devise algorithms for optimizing this objective.

In some applications, it is desirable to leave nodes unmatched if no 
suitable 
match is found. This can be achieved without having to change the problem 
formulation by adding a special ``gap'' node to the target graph for 
each node in the source graph. The gap nodes are isolated and the user has to 
define a similarity value (which can also be negative) between the source 
nodes and the gap nodes, which controls the cost of leaving a source node 
unmatched.

\begin{figure}[t]
\begin{center}
\begin{tikzpicture}[scale=0.9,every node/.style={scale=0.9}]]

\tikzstyle{action} = [thick, draw = yafcolor4!50, fill=yafcolor6!20, circle, inner sep = 1pt, text=black, minimum width=12pt]
\tikzstyle{exedge} = [black!80, thick, text=black!80]
\tikzstyle{timesegment} = [line width=4pt, draw=yafcolor4!20]
\tikzstyle{timepoint} = [draw = black!80, fill=black!80, circle, inner sep = 0pt, minimum size=6pt]
\tikzstyle{exnode1} = [thick, draw = black, fill=yafcolor4!20, circle, minimum size=4pt]
\tikzstyle{exnode2} = [thick, draw = black, fill=yafcolor5!20, circle, minimum size=4pt]
\tikzstyle{exnode3} = [thick, draw = black, fill=yafcolor6!20, circle, minimum size=4pt]

\node[exnode3, label={left:C}] (C) at (1.4, 2) {};
\node[exnode2, label={left:B}] (B) at (1, 1) {};
\node[exnode1, label={left:A}] (A) at (1.1, 0) {};

\draw[-, exedge, bend left = 10] (A) to (B);
\draw[-, exedge, bend left = 10] (B) to (C);

\node[fill=white] at (1.2,-1) {$G_s$};

\node[exnode1, label={left:$A_1$}] (A1) at (3.5, 0) {};
\node[exnode1, label={left:$A_2$}] (A2) at (5, 0) {};
\node[exnode1, label={left:$A_3$}] (A3) at (6.5, 0) {};

\node[exnode2, label={left:$B_1$}] (B1) at (4.25, 1) {};
\node[exnode2, label={left:$B_2$}] (B2) at (5.75, 1) {};

\node[exnode3, label={left:$C_1$}] (C1) at (3.5, 2) {};
\node[exnode3, label={left:$C_2$}] (C2) at (5, 2) {};

\draw[-, exedge, bend left = 10] (A1) to (B1);
\draw[-, exedge, bend right = 10] (A2) to (B1);

\draw[-, exedge, bend right = 10] (B1) to (C1);
\draw[-, exedge, bend left = 10] (B1) to (C2);

\draw[-, exedge, bend right = 10] (A3) to (B2);
\draw[-, exedge, bend right = 10] (B2) to (C2);

\node[fill=white] at (5,-1) {$G_t$};

\end{tikzpicture}
\caption{\label{figure:example-1}
An illustration of the network-alignment problem. 
The task is to align two input networks $G_s$ and $G_t$. 
Some nodes and/or edges from either network may be left unmatched. 
Letters and colors are used to indicate the sets of
candidate matching nodes,
e.g., $\Cands_{A}=\{A_1,A_2,A_3\}$, $\Cands_{B}=\{B_1,B_2\}$, and
$\Cands_{C}=\{C_1,C_2\}$.
}
\end{center}
\end{figure}
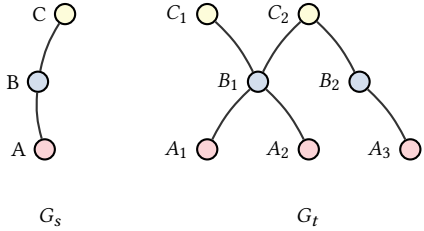

An example of the network-alignment problem is shown in Figure~\ref{figure:example-1}. 
We use letters and colors to indicate sets of matching nodes.
Indicatively, in an application of aligning social networks, 
one can think of
$A=\;${\tt Andres},
$A_1=\;${\tt Andre},
$A_2=\;${\tt Andrew}, 
$A_3=\;${\tt Andreas}, 
while
$B=\;${\tt Brendon},
$B_1=\;${\tt Brenden},
$B_2=\;${\tt Brendan}, 
etc.

\spara{Active network alignment:}
Assume now that we have access to an oracle, for example, a human expert,
with whom we can interact in the form of queries 
and obtain partial information about the correct alignment
of the two networks $G_s$ and $G_t$.

As already discussed, 
we focus on interaction
with the oracle that takes the form of the following queries: 

\smallskip
{\em Given a node $v$ in the source network $G_s$, 
and a set of candidate matches $\Cands_v$ in the target network $G_t$, 
which node from $\Cands_v$ should be matched to $v$?}

\smallskip
The oracle returns an answer $u\in \Cands_v$
and the algorithm proceeds to aligning $G_s$ 
with $G_t$ {\em given} that $v\in V_s$ is matched with $u\in V_t$.
Depending on the application, one of the candidate matches $\Cands_v$ may be the 
gap node of $v$.

The problem of active network alignment is to select the most informative
node $v\in V_s$ for which to ask the oracle to reveal
the correct matched node $u\in V_t$. 
More specifically, we aim to solve the following problem.

\begin{problem}[ActiveNetworkAlignment]
\label{problem:ana}
Given a source network $G_s=(V_s, E_s)$ and a target network $G_t=(V_t, E_t)$, select the node $v\in V_s$ for which to ask an oracle to reveal the correct matched node $u\in V_t$ so that the alignment accuracy for the remaining nodes in $V_s$ is maximized. 
\end{problem}

The {\em alignment accuracy} in Problem~\ref{problem:ana} 
is computed by fixing the alignment of $v$ to $u$, 
then aligning the remaining nodes using a standard (non-active) network alignment method, and 
finally by computing the fraction of correctly aligned unqueried nodes in $V_s$.

A natural way to approach Problem~\ref{problem:ana} is to design a function 
$\certainty\left(v\mid G_s,G_t \right)$
that quantifies the \emph{certainty}
associated with each $v\in V_s$
with respect to its match in $V_t$.
Nodes  with low certainty scores are good candidates
for being the query node,
since these nodes would otherwise be more likely to decrease the alignment accuracy. 
The active-learning framework 
provides general principles for designing such a function.
We deploy these ideas in order to design and experiment our {\certainty} function.
More details on this are given in Sections~\ref{section:algorithms} and 
\ref{section:experiments}.

In the example of Figure~\ref{figure:example-1},
we see that $A$ in $G_s$ is most similar to $A_1$, $A_2$, and $A_3$ in $G_t$.
If $A$ is matched to $A_1$ or $A_2$, then $B$ should be uniquely matched to $B_1$.
If $A$ is matched to $A_3$, then $B$ should be uniquely matched to $B_2$ and $C$ to $C_2$. 
Thus, matching node $A$ first, to a large extent, determines the rest of the alignment.
On the other hand, matching $B$ or $C$ first
does not determine the rest of the alignment with the same level of certainty.
We conclude that, in this example, 
it is a good strategy to ask the oracle to provide us with the correct alignment for node~$A$.

\section{Matching-based active network alignment}
\label{section:algorithms}

In this section, we present the proposed strategy for active network alignment.
Our strategy identifies which node $v\in V_s$ to query 
via a novel approach for quantifying the certainty of each node. 
The main idea is as follows:
instead of solving the alignment problem to find only a single matching (the optimal network alignment),
we sample a number of high-quality matchings (near-optimal network alignments), 
and then, compute the certainty of each node by considering 
the distribution of its matched nodes on the sampled matchings.

We study two different methods for sampling matchings;  
\gm, where matchings are sampled according to their score
(so better matchings have higher probability of being included in the sample), and 
\tm, where the top-$\ell$ matchings are considered.
We also consider different alternatives for computing node certainty 
based on the sampled matchings. 

We now discuss our approach in more detail. 
We start by describing our strategy at a high level, 
and then proceed to the description of each of its components
and different alternatives.

\subsection{Overview}

The general active network alignment approach, in which nodes are queried iteratively,
can be summarized as follows:
In every iteration pick a node $\hat{v}\in V_s$. 
The node $\hat{v}$ together with 
its set of candidate matching nodes $\Cands_{\hat{v}}\subseteq V_t$  
is shown to the (human) oracle, and
the oracle selects a node $\hat{u}\in \Cands_{\hat{v}}$
as the best match for $\hat{v}$. 
Assert that $\hat{u}$ is the best matching for $\hat{v}$
by updating the candidate node sets 
so that $\Cands_{\hat{v}}=\{\hat{u}\}$, 
and removing $\hat{u}$ from all other sets of candidate nodes.
When the budget of oracle queries is exhausted, 
solve the remaining network alignment problem to align the unqueried nodes.
Pseudocode for this approach is shown in Algorithm~\ref{algo:generalq} (different steps of the algorithm are explained later in this section).

\begin{algorithm}[t]
  \begin{algorithmic}[1]
      \Statex {\bf Input:} $G_s=(V_s,E_s)$, $G_t=(V_t,E_t)$, candidate 
      matches \Cands, and query budget $k$.
      \Statex {\bf Output:} An alignment between $V_s$ and $V_t$. 
      \For{$i=1,\ldots ,k$}
        \State $H = $ \alignment
        \State $\hat{v} =\argmin_{v\in V_s}\certainty\left(v\mid H,G_s,G_t 
        \right)$ \label{line:score}
        \State $\hat{u}=\oracleq(\hat{v},\Cands_{\hat{v}})$, with $\hat{u}\in 
        \Cands_{\hat{v}}$
        \State $\Cands_{\hat{v}} = \{\hat{u}\}$ \Comment{Update candidate 
        matches.}
        \For{$v \in V_s \setminus \{\hat{v}\}$}
        \State $\Cands_v = \Cands_v \setminus \{\hat{u}\}$
        \EndFor
      \EndFor
      \State $H = $ \alignment
      \State {\bf return} \bm{$H$}
  \end{algorithmic}
  \caption{\label{algo:generalq} The general active network alignment framework.}
\end{algorithm}

Our main contribution, 
is the methodology for quantifying the certainty of the candidate
nodes to be queried at every iteration of Algorithm~\ref{algo:generalq}. 
Our approach consists of the following steps.

\squishlist
\item[{\bf Step 1.}]  
Construct a weighted bipartite graph 
$H=(V_s, V_t,E_h)$ that forms the basis
for the matching-based network-alignment algorithm. 
\item[{\bf Step 2.}] 
Sample a set $\calM_\ell$ of $\ell$ high-quality matchings in~$H$.
\item[{\bf Step 3.}]
For each node $v\in V_s$, estimate the \emph{certainty} we have about the 
correct match for $v$.
These certainty values, $\certainty(v)$, are computed using the information in 
the set of sampled matchings~$\calM_\ell$.
\item[{\bf Step 4.}]  
Identify the node 
$\hat{v}\in V_s$ with the {\em least certainty}, 
and query the oracle for selecting the best match of $\hat{v}$ 
among the set of candidate matching nodes $\Cands_{\hat{v}}\subseteq V_t$.
\squishend

For sampling a set of matchings $\calM_\ell$ 
we consider two alternatives: 
\gm (sample matchings according to their score) and 
\tm (take the top-$\ell$ matchings); 
those are discussed in Section~\ref{section:sampling-matchings}.
We also consider three different ways of estimating node certainty ({\bf Step 3}) 
presented in Section~\ref{section:quantifying-certainty}.
First we discuss how to construct the weighted bipartite graph $H=(V_s, V_t,E_h)$ 
on which the set of sampled matchings $\calM_\ell$ is obtained.

\subsection{Constructing the bipartite graph $\boldsymbol{H}$}
\label{section:natalie-netalign}
Our strategy is applicable to 
any {\sc NetworkAlignment} algorithm that is based on 
solving a bipartite-matching problem. 
In our experiments, we use 
{\natalie}~\cite{el2015} and {\netalign}~\cite{bayati2013}, 
two state-of-the-art algorithms.
Both methods aim to solve the following quadratic integer program, adopting, 
however, different approaches.

\smallskip
\noindent
{\em Integer-programming formulation:}
The IP formulation of {\natalie} and {\netalign}
introduces a variable $x_{ij}$ for each pair of nodes $i\in V_s$ and $j\in V_t$.
The variable $x_{ij}$ is set to 1 if $i$ is matched to $j$, 
while it is set to 0 otherwise. 
We also use $\neighbors(v)$, for $v\in V_s \cup V_t$, 
to denote the set of all pairs $\{(i,j)\}$ for which $v=i$ or $v=j$,
and $A_{s;ik}$ ($A_{t;ik}$) to denote whether there is an edge between nodes $i$ and $k$ in the source (target) graph.
\begin{align*}
 \max_{x}  \quad
 & \sum_{(i,j) \in V_s \times V_t} \sigma(i,j) x_{ij} \\ & 
 + g \sum_{(i,j) \in 
V_s \times V_t} \sum_{(k,\ell) \in V_s \times V_t} 
A_{s;ik} A_{t;j\ell} x_{ij} x_{k\ell}, \\
\textmd{such that}\quad 
 & \sum_{(i,j)\in\neighbors(v)} x_{ij} \leq 1, \quad \textmd{for all } v \in V_s\cup V_t, \\
 & x_{ij} \in \{0,1\}, \quad \textmd{for all } (i,j) \in V_s \times V_t.
\end{align*}
In the above integer program, 
each pair of matched nodes $i$ and $j$ contributes a reward of value $\similarity(i,j)$, 
while an additional reward of value $g$ is given 
if an edge in $E_s$ is matched to an edge in $E_t$.
The user-defined parameter $g$ quantifies
the relative importance between correctly-matched nodes and correctly-matched 
edges, and it is typically set based on prior knowledge or 
cross-validation.
The inequality constraint ensures that the solution is a matching.

\smallskip
\noindent
{\em Solution via Lagrangian relaxation:}
\natalie, originally proposed by Klau~\cite{klau2009}, solves the quadratic 
integer program by first 
linearizing it and then employing a Lagrangian relaxation technique. Klau shows 
that the original integer program is \np-hard, but remarkably, by 
relaxing a symmetry constraint for the linearized quadratic terms, the problem 
becomes solvable in polynomial time via multiple maximum-weight bipartite matchings.

\natalie iteratively updates the Lagrangian multipliers $\lambda$ for the 
relaxed constraints using subgradient optimization. 
The solutions of the 
relaxed problem 
provide upper bounds for the 
original problem, whereas the feasible solutions, which can be directly 
extracted from the relaxed solutions, provide lower bounds.

The best feasible solution found provides
the bipartite graph $H$ for our active strategy.

\smallskip
\noindent
{\em Solution via message passing:} The \netalign algorithm, proposed by Bayati 
et al.~\cite{bayati2013}, solves the same optimization problem using a belief 
propagation (BP) approach. 
This approach makes local, greedy updates 
by passing messages between the neighboring nodes. To obtain an 
integral solution, \netalign constructs and solves a maximum-weight matching 
problem based on the BP messages at every iteration of the algorithm.

Again, we set $H$ to correspond to the matching problem that gives the 
best solution.

\smallskip
Due to space constraints, we do not discuss in detail how the  
weights of the edges of the bipartite graphs are set. 
Note, however, that the weights aim at capturing both 
the feature-based and structural similarities of the matching nodes, 
and the higher the weight the more similar the nodes are.
For more details we refer the reader to the original papers~\cite{bayati2013,el2015}.

\subsection{Sampling matchings}
\label{section:sampling-matchings}

Next we present two approaches for sampling matchings: \gm, which first defines a probability distribution over the space of matchings and then employs Gibbs sampling to draw samples from this space, and \tm, which computes the top-$\ell$ matchings.

\spara{Gibbs sampling for matchings:} Markov chain Monte Carlo 
techniques are popular for drawing samples from complex multi-di\-men\-sional distributions \cite{bishop2009}. 
In order to apply these methods to sample matchings $M$, we need to define a probability distribution over the space of matchings induced by the bipartite graph $H$. 
Similar to Volkovs and Zemel~\cite{volkovs2012}, we use the standard Gibbs form
\begin{align}
 \prob{M \mid H} &= \frac{1}{Z(M,\beta)} \exp\left(-\frac1\beta E(M,H)\right) \nonumber \\
 &= \frac{1}{Z(M,\beta)} \exp\left(\frac1\beta \sum_{v \in V_s} H_{v,M(v)}\right),
\end{align}
where $M(v)$ denotes the node to which $v$ is matched to in $M$, 
$Z(M,\beta)$ is the partition function that normalizes the distribution, and 
$\beta$ is a ``temperature'' constant that defines the smoothness of the distribution. 
The value of $\beta$ is optimized using a training dataset, 
as discussed in Section~\ref{sec:query_comp}.

We sample $\ell$ matchings $\calM_\ell=\{M_1,\ldots ,M_\ell\}$ from this distribution as follows: we initialize $M$ 
with the maximum-weight matching. Then 
at each iteration $i$, we go through the nodes in $V_s$ in a random 
order and for each node $v \in V_s$, we pick one of the candidate matching 
nodes $u 
\in \Cands_{v}$ uniformly at random. Next we consider re-assigning $v$ to $u$ 
and $v'$ to $M(v)$ where $v'$ is the node currently matched to $u$ if any. If 
$M(v)$ is not among the candidate matches of $v'$, we pick another node $u$ 
uniformly at random among the remaining candidates of $v$ until a possible 
re-assignment has been found 
($v$ can always be re-assigned to its current  match $M(v)$).
The new matching $M'$, which would result from the re-assignment, is 
realized and the update $M \leftarrow M'$ performed with probability
\begin{small}
\begin{align}
&\prob{M' \mid M, H} = \frac{\exp\left(-\frac1\beta E(M',H)\right)}{\exp\left(-\frac1\beta E(M,H)\right) + \exp\left(-\frac1\beta E(M',H)\right)} \nonumber \\
 &= \frac{\exp\left(\frac1\beta \left(H_{v,u} + H_{v',M(v)}\right)\right)}{\exp\left(\frac1\beta 
 \left(H_{v,M(v)} + H_{v',u}\right)\right) + \exp\left(\frac1\beta \left(H_{v,u} + 
 H_{v',M(v)}\right)\right)}.
\end{align}
\end{small}
Once each node has been processed, we let $i$-th sample be $M_i \leftarrow M$. 
Then, we generate a new random permutation of the nodes $V_s$ and 
repeat the process until $\ell$ samples have been drawn. 
We call this method \gm.\footnote{A limitation of this swap-based approach is that if the problem at hand has nodes with partially overlapping sets of candidate matches, there will be some states that are unreachable by the Markov chain (i.e. the chain is non-ergodic).}
Assuming that each node has $c$ candidate matches, the worst-case
complexity of \gm is $\mathcal{O}(\ell c |V_s|)$.

Volkovs and Zemel \cite{volkovs2012} note that Gibbs sampling is often found to 
mix slowly and get trapped in local modes. 
In our case, 
such slow mixing implies that we will end up 
exploring only the areas around the maximum weight matching, 
which is the starting point of the sampling.
However, even if this happens, 
we do not expect it to be a problem 
as we indeed are interested in sampling high-quality matchings
close to the optimal solution.\footnote{We also tested the 
sequential matching sampler proposed by Volkovs and Zemel~\cite{volkovs2012} 
which outperformed a Gibbs sampler in their experiments. However, in our case, 
this approach suffered from extremely low acceptance rates, even after 
adjusting parameter $\rho$ \cite{volkovs2012}. This might be related to the 
distribution of the weights in $H$ or to the small sizes of the graphs studied 
in \cite{volkovs2012}.}

\spara{Top-$\ell$ maximum-weight matchings:}
To compute the top-$\ell$ max\-imum-weight matchings $\calM_\ell$,
we use an algorithm invented by 
Murty~\cite{murty68algorithm} in 1968.
This algorithm first computes the best maximum-weight matching in $H$, then 
splits the problem into $\mathcal{O}(n)$ smaller matching problems and 
reconstructs the second best matching based on the solutions of the smaller 
problems. The process is repeated until $\ell$ matchings have been discovered, 
and in total, the algorithm has a running time of
$\mathcal{O}(\ell n^4)$, 
where $n$ is the total number of nodes in the graph, 
i.e., in our case $n=\max\{|V_s|,|V_t|\}$. Using a modification by Miller et 
al.~\cite{miller1997}, the algorithm runs in time $\mathcal{O}(\ell n^3)$.

\begin{figure*}[t]
\begin{center}
\begin{tikzpicture}[scale=0.9,every node/.style={scale=0.9}]]

\tikzstyle{action} = [thick, draw = yafcolor4!50, fill=yafcolor6!20, circle, inner sep = 1pt, text=black, minimum width=12pt]
\tikzstyle{exedge} = [black!80, line width=0.45mm, text=black!80]
\tikzstyle{backedge} = [black!30, dashed, line width=0.20mm, text=black!80]
\tikzstyle{timesegment} = [line width=4pt, draw=yafcolor4!20]
\tikzstyle{timepoint} = [draw = black!80, fill=black!80, circle, inner sep = 0pt, minimum size=6pt]
\tikzstyle{exnode1} = [thick, draw = black, fill=yafcolor4!20, circle, minimum size=4pt]
\tikzstyle{exnode2} = [thick, draw = black, fill=yafcolor5!20, circle, minimum size=4pt]
\tikzstyle{exnode3} = [thick, draw = black, fill=yafcolor6!20, circle, minimum size=4pt]


\node[exnode1, label={left:A}] (A) at (-2.5, 0) {};
\node[exnode2, label={left:B}] (B) at (-2.5, 1) {};
\node[exnode3, label={left:C}] (C) at (-2.5, 2) {};

\node[exnode1, label={right:$A_1$}] (A1) at (-1.5, -0.2) {};
\node[exnode1, label={right:$A_2$}] (A2) at (-1.5, 0.2) {};
\node[exnode1, label={right:$A_3$}] (A3) at (-1.5, 0.6) {};
\node[exnode2, label={right:$B_1$}] (B1) at (-1.5, 1.0) {};
\node[exnode2, label={right:$B_2$}] (B2) at (-1.5, 1.4) {};
\node[exnode3, label={right:$C_1$}] (C1) at (-1.5, 1.8) {};
\node[exnode3, label={right:$C_2$}] (C2) at (-1.5, 2.2) {};

\node[fill=white] at (-2,3.5) {Bipartite graph};
\node[fill=white] at (-2,3) {used for matching};

\draw[-, backedge] (A) to (A1);
\draw[-, backedge] (A) to (A2);
\draw[-, backedge] (A) to (A3);
\draw[-, backedge] (B) to (B1);
\draw[-, backedge] (B) to (B2);
\draw[-, backedge] (C) to (C1);
\draw[-, backedge] (C) to (C2);


\node[exnode1, label={left:A}] (Am1) at (1, 0) {};
\node[exnode2, label={left:B}] (Bm1) at (1, 1) {};
\node[exnode3, label={left:C}] (Cm1) at (1, 2) {};

\node[exnode1, label={right:$A_1$}] (A1m1) at (2, -0.2) {};
\node[exnode1, label={right:$A_2$}] (A2m1) at (2, 0.2) {};
\node[exnode1, label={right:$A_3$}] (A3m1) at (2, 0.6) {};
\node[exnode2, label={right:$B_1$}] (B1m1) at (2, 1.0) {};
\node[exnode2, label={right:$B_2$}] (B2m1) at (2, 1.4) {};
\node[exnode3, label={right:$C_1$}] (C1m1) at (2, 1.8) {};
\node[exnode3, label={right:$C_2$}] (C2m1) at (2, 2.2) {};

\draw[-, backedge] (Am1) to (A1m1);
\draw[-, backedge] (Am1) to (A2m1);
\draw[-, backedge] (Am1) to (A3m1);
\draw[-, backedge] (Bm1) to (B1m1);
\draw[-, backedge] (Bm1) to (B2m1);
\draw[-, backedge] (Cm1) to (C1m1);
\draw[-, backedge] (Cm1) to (C2m1);

\draw[-, exedge] (Am1) to (A1m1);
\draw[-, exedge] (Bm1) to (B1m1);
\draw[-, exedge] (Cm1) to (C1m1);


\node[exnode1, label={left:A}] (Am2) at (4, 0) {};
\node[exnode2, label={left:B}] (Bm2) at (4, 1) {};
\node[exnode3, label={left:C}] (Cm2) at (4, 2) {};

\node[exnode1, label={right:$A_1$}] (A1m2) at (5, -0.2) {};
\node[exnode1, label={right:$A_2$}] (A2m2) at (5, 0.2) {};
\node[exnode1, label={right:$A_3$}] (A3m2) at (5, 0.6) {};
\node[exnode2, label={right:$B_1$}] (B1m2) at (5, 1.0) {};
\node[exnode2, label={right:$B_2$}] (B2m2) at (5, 1.4) {};
\node[exnode3, label={right:$C_1$}] (C1m2) at (5, 1.8) {};
\node[exnode3, label={right:$C_2$}] (C2m2) at (5, 2.2) {};

\draw[-, backedge] (Am2) to (A1m2);
\draw[-, backedge] (Am2) to (A2m2);
\draw[-, backedge] (Am2) to (A3m2);
\draw[-, backedge] (Bm2) to (B1m2);
\draw[-, backedge] (Bm2) to (B2m2);
\draw[-, backedge] (Cm2) to (C1m2);
\draw[-, backedge] (Cm2) to (C2m2);

\draw[-, exedge] (Am2) to (A2m2);
\draw[-, exedge] (Bm2) to (B1m2);
\draw[-, exedge] (Cm2) to (C1m2);


\node[exnode1, label={left:A}] (Am3) at (7, 0) {};
\node[exnode2, label={left:B}] (Bm3) at (7, 1) {};
\node[exnode3, label={left:C}] (Cm3) at (7, 2) {};

\node[exnode1, label={right:$A_1$}] (A1m3) at (8, -0.2) {};
\node[exnode1, label={right:$A_2$}] (A2m3) at (8, 0.2) {};
\node[exnode1, label={right:$A_3$}] (A3m3) at (8, 0.6) {};
\node[exnode2, label={right:$B_1$}] (B1m3) at (8, 1.0) {};
\node[exnode2, label={right:$B_2$}] (B2m3) at (8, 1.4) {};
\node[exnode3, label={right:$C_1$}] (C1m3) at (8, 1.8) {};
\node[exnode3, label={right:$C_2$}] (C2m3) at (8, 2.2) {};

\draw[-, backedge] (Am3) to (A1m3);
\draw[-, backedge] (Am3) to (A2m3);
\draw[-, backedge] (Am3) to (A3m3);
\draw[-, backedge] (Bm3) to (B1m3);
\draw[-, backedge] (Bm3) to (B2m3);
\draw[-, backedge] (Cm3) to (C1m3);
\draw[-, backedge] (Cm3) to (C2m3);

\draw[-, exedge] (Am3) to (A1m3);
\draw[-, exedge] (Bm3) to (B1m3);
\draw[-, exedge] (Cm3) to (C2m3);


\node[exnode1, label={left:A}] (Am4) at (10, 0) {};
\node[exnode2, label={left:B}] (Bm4) at (10, 1) {};
\node[exnode3, label={left:C}] (Cm4) at (10, 2) {};

\node[exnode1, label={right:$A_1$}] (A1m4) at (11, -0.2) {};
\node[exnode1, label={right:$A_2$}] (A2m4) at (11, 0.2) {};
\node[exnode1, label={right:$A_3$}] (A3m4) at (11, 0.6) {};
\node[exnode2, label={right:$B_1$}] (B1m4) at (11, 1.0) {};
\node[exnode2, label={right:$B_2$}] (B2m4) at (11, 1.4) {};
\node[exnode3, label={right:$C_1$}] (C1m4) at (11, 1.8) {};
\node[exnode3, label={right:$C_2$}] (C2m4) at (11, 2.2) {};

\draw[-, backedge] (Am4) to (A1m4);
\draw[-, backedge] (Am4) to (A2m4);
\draw[-, backedge] (Am4) to (A3m4);
\draw[-, backedge] (Bm4) to (B1m4);
\draw[-, backedge] (Bm4) to (B2m4);
\draw[-, backedge] (Cm4) to (C1m4);
\draw[-, backedge] (Cm4) to (C2m4);

\draw[-, exedge] (Am4) to (A2m4);
\draw[-, exedge] (Bm4) to (B1m4);
\draw[-, exedge] (Cm4) to (C2m4);


\node[exnode1, label={left:A}] (Am5) at (13, 0) {};
\node[exnode2, label={left:B}] (Bm5) at (13, 1) {};
\node[exnode3, label={left:C}] (Cm5) at (13, 2) {};

\node[exnode1, label={right:$A_1$}] (A1m5) at (14, -0.2) {};
\node[exnode1, label={right:$A_2$}] (A2m5) at (14, 0.2) {};
\node[exnode1, label={right:$A_3$}] (A3m5) at (14, 0.6) {};
\node[exnode2, label={right:$B_1$}] (B1m5) at (14, 1.0) {};
\node[exnode2, label={right:$B_2$}] (B2m5) at (14, 1.4) {};
\node[exnode3, label={right:$C_1$}] (C1m5) at (14, 1.8) {};
\node[exnode3, label={right:$C_2$}] (C2m5) at (14, 2.2) {};

\draw[-, backedge] (Am5) to (A1m5);
\draw[-, backedge] (Am5) to (A2m5);
\draw[-, backedge] (Am5) to (A3m5);
\draw[-, backedge] (Bm5) to (B1m5);
\draw[-, backedge] (Bm5) to (B2m5);
\draw[-, backedge] (Cm5) to (C1m5);
\draw[-, backedge] (Cm5) to (C2m5);

\draw[-, exedge] (Am5) to (A3m5);
\draw[-, exedge] (Bm5) to (B2m5);
\draw[-, exedge] (Cm5) to (C2m5);

\draw [decorate,decoration={brace,amplitude=10pt}]
(0.5,3) -- (14.5,3) node [black,midway,yshift=0.6cm] {A sample of five matchings};

\end{tikzpicture}
\caption{\label{figure:example-2}
The instance of the network-alignment problem in Figure~\ref{figure:example-1}
is transformed to an instance of a maximum-weight matching problem.
Computing the set $\calM_\ell$ of $\ell=5$ matchings helps to quantify
the uncertainty of a node in network alignment.
}
\end{center}
\end{figure*}
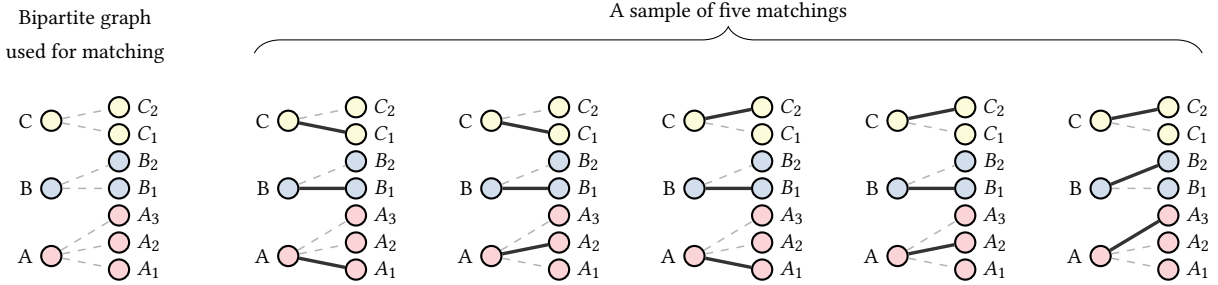

\subsection{Quantifying certainty and identifying the node to query}
\label{section:quantifying-certainty}

Using uncertainty to determine which data points should be labeled
is a standard active-learning strategy~\cite{settles2010}. 
The idea is to select for labeling the data points for which we are the 
{\em least certain} 
as to what the correct output should be.

A natural approach for quantifying the uncertainty of a match $M(v)$ of node $v \in V_s$ 
is to consider the marginal distribution $\prob{M(v)=u \mid H}$, where $u \in V_t$. 
Given the set of sampled matchings $\calM_\ell$, 
the marginal distribution can be obtained simply by computing the fraction of samples for which node $v$ is matched to $u$. We then define the {\em certainty} of node $v$ as  
\begin{equation}
\label{eq:certainty}
\certainty(v) = \max_{u\in C_v} \prob{M(v)=u \mid H}.
\end{equation}
The intuition for Equation~(\ref{eq:certainty})
is that when a node $v$ is matched to the same node $u$ in most of the matchings
in $\calM_\ell$, then there is little uncertainty; 
based on the evidence in the set of sampled matchings $\calM_\ell$, 
$u$ is a good match for $v$
and no extra information will be gained by querying $v$. 
On the other hand, if the evidence in $\calM_\ell$ is inconclusive with respect to the best match for
$v$, then $v$ should be queried.  
Accordingly, the proposed strategy is to query the node $\hat{v}$ with the least certainty, i.e., 
\begin{equation}
\label{eq:topkcriterion}
\hat{v} = \argmin_{v\in V_s} \certainty(v). 
\end{equation}

To obtain further intuition for the proposed method, 
consider again the example in Figure~\ref{figure:example-1}.
The bipartite graph $H$ for this example 
is shown on the left in Figure~\ref{figure:example-2}.

Now assume that we draw a sample of $\ell$ matchings in this bipartite graph.
A sample of five matchings is shown in Figure~\ref{figure:example-2}.
For each node $v\in V_s$ 
we can compute the distribution of the nodes in $V_t$ 
that $v$ is matched to in these five matchings. 
A node with high uncertainty is a node that is matched to many different 
nodes in $V_t$ and the distribution of matches is well-balanced.
In the example of Figure~\ref{figure:example-2},
the distribution of matchings for nodes $A$, $B$, and $C$ is 
$\{A_1:\p{40},\,A_2:\p{40},\,A_3:\p{20}\}$, 
$\{B_1:\p{80},\,B_2:\p{20}\}$, and  
$\{C_1:\p{40},\,C_2:\p{60}\}$, respectively.
From these distributions
one can argue that node $A$ has the highest uncertainty. 

An alternative measure for the certainty of a node would be to study entropy, 
which is commonly used in active learning~\cite{settles2008analysis,roy2001toward}.
However, in our experiments we found that 
Equation~\eqref{eq:certainty} consistently outperformed a method which defined 
certainty as the negative entropy of the marginal distribution. Thus we do 
not report experimental results using the entropy-based method.

A limitation of Equation~\eqref{eq:certainty} is that it measures certainty only 
locally and thus it does not capture the effect that knowing the matching for a 
given node $v$ would have on the remaining alignment. Therefore, we consider 
also another alternative query strategy which 
queries the node $\hat{v}$ 
such that once $\hat{v}$ is queried and its match is known, the expected
certainty 
of the remaining nodes is maximized. That is, 
\begin{equation}
\label{eq:criterion2}
\hat{v} = \arg \max_{v\in V_s} \sum_{u \in C_v} \left(\prob{M(v)=u \mid H} 
\sum_{v' \in V_s \setminus v} \certainty(v')\right). 
\end{equation}
This approach is inspired by previous works in sensor 
placement~\cite{krause2008,krause2007near} and active learning~\cite{macskassy2009}. 
As pointed out by the previous works, a limitation 
of this approach is its high computational complexity, which in our case is 
given by $\mathcal{O}(c^2 |V_s|^2)$, assuming that each node has $c$ candidate 
matches. In our initial experiments, it performed similarly to the simpler 
query strategy given in Equation~\eqref{eq:topkcriterion}, which is why we only 
consider the latter hereafter.

\subsection{Batch querying} 
\label{sec:batch}

Instead of finding the best query to ask at every 
iteration of Algorithm~\ref{algo:generalq} we can
adopt a ``batch'' approach, where in each iteration we identify a batch of 
queries to ask --- the ones
that correspond to the nodes with the smallest $\certainty$ values.
When 
considering batches of size
$k'<k$, where $k$ is the total number of queries to be made, we can achieve a 
speedup of the order $\lceil \frac{k}{k'} \rceil$. 

We can readily employ all the query strategies studied in this paper to query 
in batches. However, these strategies might yield 
nodes that strongly depend on each other (that is, if we knew the alignment for 
one node, we could easily align another node). Thus, developing more 
sophisticated methods for batch querying remains an interesting avenue for 
future research.

\section{Experimental evaluation}
\label{section:experiments}

We evaluate a number of different network-alignment query strategies
on networks where the correct alignment is known. 
Our methodology is as follows: 
First, we solve the network-alignment problem 
with a non-active alignment method and report the alignment accuracy, 
that is, the fraction of correctly aligned nodes. 
Then we start simulating the oracle queries by fixing the nodes to their correct matches. 
After each query
we solve the network-alignment problem given the current correct matches
and report the alignment accuracy on the unqueried nodes. 
This process is repeated for each query strategy separately with the same initial graphs. 
A good strategy should use 
as few queries as possible to reach the desired alignment accuracy.

As mentioned earlier, the query strategies discussed in this paper can be 
applied on top of any (non-active) network-alignment method 
that results in solving a bipartite-matching problem. 
In the experiments, we employ the query 
strategies on top of two state-of-the-art alignment methods, 
\natalie  \cite{el2015,klau2009} and \netalign \cite{bayati2013} (see  Section~\ref{section:natalie-netalign}
for a brief discussion).
\edit{The query strategies as well as the non-active network 
alignment methods 
used in the experiments support leaving nodes unmatched, but in our 
experiments, there is always a match for each node of the source graph.}

Next we describe our datasets and present an empirical comparison between \gm, 
\tm, and four baseline query strategies.

\subsection{Datasets}
\label{sec:data}
We use three types of datasets. 
In all datasets,
there is a sufficient degree of ambiguity in the node attributes
so that for each node in one graph there are 
several candidate matches in the other graph. 
Furthermore, the edges of the graphs may have been corrupted.

\spara{Preferential-attachment graphs:}
We generate a network using the preferential-attachment 
model \cite{barabasi1999}, which captures some of the key characteristics 
of social networks and has been previously used to study network-alignment 
methods~\cite{korula2014}. The network consists of $1\,000$ nodes and 2 edges per 
new node. For each node, we sample a label from a set of 33 
unique labels and treat the labels as attributes so that the similarity between 
two nodes is set to 1 if their labels are the same and 0 otherwise. Only the 
nodes with the same label are considered candidate matches, resulting in 
30 candidate matches per node on average.

We make two copies of the network and in each copy 
we independently corrupt
the edges to further complicate the alignment task. 
We first discard $60\,$\% of the graph edges chosen at random
and then add $50\,$\% more edges again selected at random.

\spara{Social networks:}
We use the the multiplex dataset from Aarhus University \cite{magnani2013}, 
which contains five networks between the employees of the Department of Computer Science. 
We select two of these layers, the Facebook and the lunch networks, and try to align 
the former to the latter. 
Since the dataset only contains anonymized user identifiers, we again 
sample a label for each node so that on average there are three people with the 
same label. The lunch network contains 60 people, whereas the Facebook network 
covers only a subset of 32 of them. While this dataset is small, 
it makes an interesting case study since it contains two very different 
types of real-world networks and the correct alignment is known.

\spara{Family trees:}
Aligning family trees is an import problem encountered in various online 
genealogy services where different people provide family tree fragments, 
which then, ideally, get merged into a single large family tree.
We have obtained a family tree\footnote{Note that in the graph-theoretic sense 
family trees are not trees since they contain cycles such as 
\textit{Mother--FirstChild--Father--SecondChild--Mother}.} containing $64\,208$ 
people constructed by an individual genealogical researcher. 
We sample a subgraph of this network by randomly picking a 
seed person and doing a random walk until $1\,000$ distinct people have been 
discovered.

For each person we only consider the first name, last name, and birth year.
Birth year is 
corrupted by rounding it to the nearest ten. Some of the first and last names 
are randomly replaced by their alternative spellings based on a list of 
common name variations. For example, the name \textit{Felix Ahlrooth} might get 
replaced by \textit{Feeliks Alroot}. Such name variations are 
frequently encountered in historical documents used in genealogical research. 
From both graphs independently, we 
discard \p{50} of the edges at random.

The task is to align the subgraph into the larger graph. When selecting 
candidate entities for each in\-di\-vi\-dual in the subgraph, 
we find people from the larger graph born in the same decade and select four of 
them with the most 
similar names in addition to the groundtruth match. Name similarity is 
computed as the average Jaro-Winkler similarity \cite{winkler1990} 
between the first names and between the last names.\footnote{The Jaro-Winkler similarity is a popular choice for 
de-duplicating name records.}

\subsection{Baseline query strategies}
We compare the performance of the sampling-based query strategies to 
four baseline strategies.

\spara{\margin}: Querying data samples with a small score difference 
(\textit{margin}) between the two most probable labels is a common 
approach in active 
learning \cite{settles2010}. In the context of active network alignment, this 
method has been previously used by Cort\'{e}s and Serratosa 
\cite{cortes2013}.

Thus, given the bipartite graph $H$, {\margin} computes 
$\certainty(v)$ for every $v\in V_s$ by first
finding the two edges incident to $v$ with the largest weights $w_1(v)$ 
and $w_2(v)$ in $H$. Then the node is scored according to
\[
\certainty(v) = w_1(v)-w_2(v).
\]
Intuitively, the larger the difference between $w_1(v)$ and $w_2(v)$, the less 
uncertainty there is with respect to the best match for node $v$. 

\spara{\lccl}: The \textit{Least  Confident  given  the  Current  Labelling} 
query strategy ranks nodes according to
\[
 \certainty(v) = H_{v,M(v)},
\]
where $M$ is the current matching (the maximum-weight matching in $H$) and 
$H_{v,M(v)}$ is the weight between node $v$ and its current match. This 
method was reported to yield the highest precision of the four uncertainty 
sampling based methods studied by Cort\'{e}s and Serratosa \cite{cortes2013}. 

\spara{\betw}: This method queries a previously unqueried node with the highest 
betweenness centrality in $G_s$. The method has been previously used by 
Macskassy \cite{macskassy2009}.

\spara{\random}: This method queries nodes from $V_s$ in a random order.

\begin{figure*}[!t]
	\centering
	\includegraphics[width=0.99\textwidth]{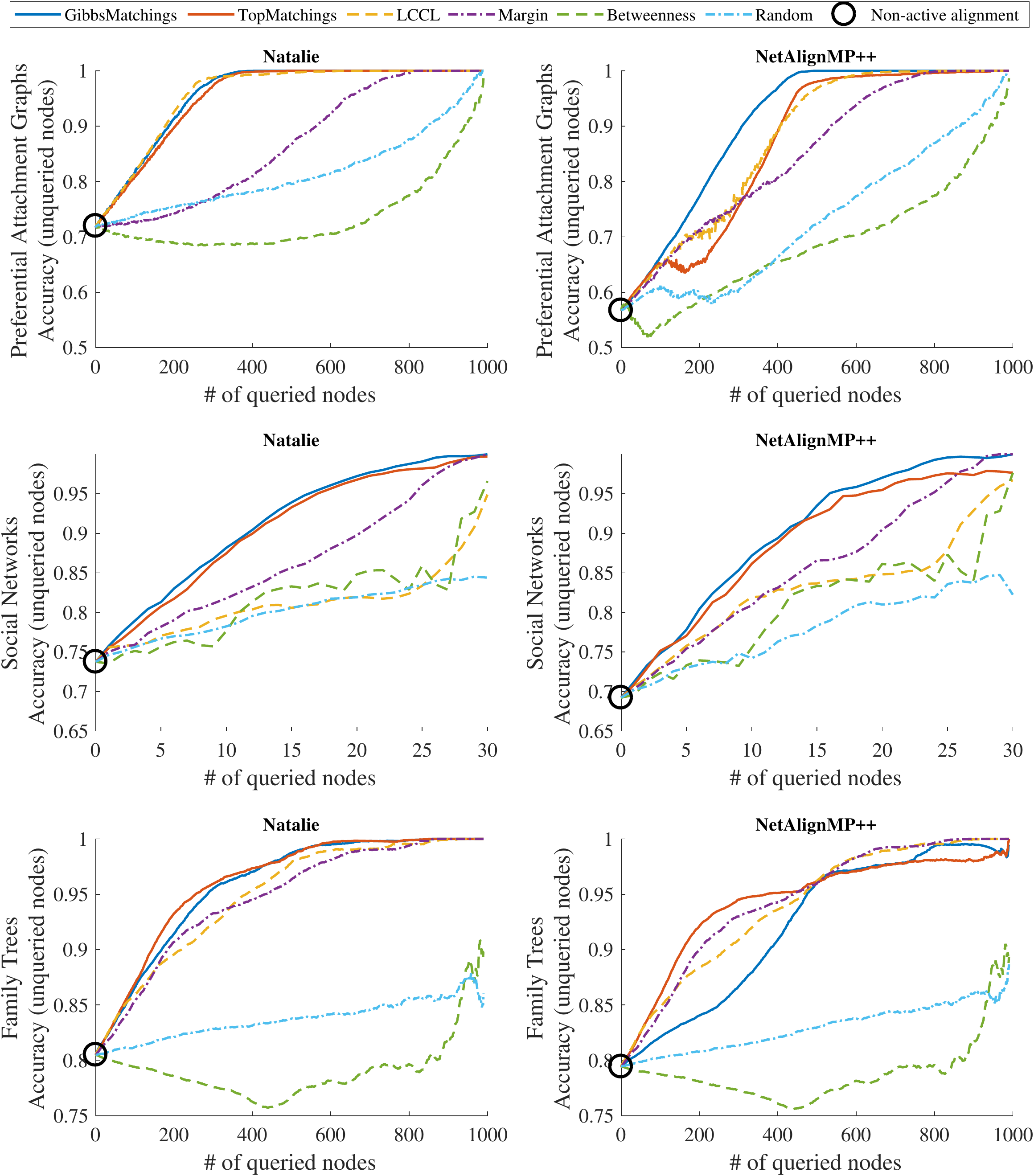}
	\caption{Alignment accuracy of the unqueried nodes for three datasets: preferential attachment graphs (\textit{top}), social networks (\textit{middle}), and family trees (\textit{bottom}). In most cases, the proposed methods (\gm and \tm) achieve a high accuracy with fewer queries than the four baseline methods. Furthermore, the proposed active-querying strategies clearly improve the initial solution obtained by a non-active alignment method.}
	\label{fig:res1}
\end{figure*}

\subsection{Query strategy comparison}
\label{sec:query_comp}

Here we perform an empirical comparison of the different query strategies. For 
each dataset, we run a minimum of 30 random initializations of the input graphs
and average the results; for the preferential-attachment graphs, we generate a 
new pair of input graphs at each initialization, for social networks, we sample 
new node labelings and for family trees a new subgraph at each initialization.

\spara{Adjusting hyperparameters:}
The number of matchings $\ell$ in \tm is set to 30. We 
also tried larger values up to 3000 but this did not seem to have a significant 
effect on the results. In \gm, we set the number of samples $\ell$ to 3000. Additionally, we have to choose the value of the temperature parameter $\beta$. If the value is set too high, the distribution becomes close to a uniform distribution, whereas if it is set too low, the distribution becomes very concentrated and the swaps to a lower energy state will always fail. Therefore, we first normalize the bipartite graph $H$ by dividing its values with the difference of the average minimum and maximum value of the different rows of $H$.
Then we optimize $\beta$ using a separate training dataset. This dataset is generated using the preferential-attachment model, 
as described in Section~\ref{sec:data} but with only 50 nodes per graph and 5 unique labels. Among the values we tried, $\{0.001, 0.01, 0.1, 1\}$, we found $\beta = 0.1$ to produce the most accurate predictions for the training data, and thus we fix this value for the rest of the experiments.

\spara{Results:}
The alignment accuracies are shown in Figure \ref{fig:res1}. 
On the $x$-axis, we have the number of 
queried nodes and on the $y$-axis, the alignment accuracy for the unqueried 
nodes, that is, the fraction of correctly aligned unqueried nodes.

In the top row of Figure \ref{fig:res1}, we see that by querying the 400 best nodes 
according to \gm, \tm, or \lccl, the remaining 600 nodes get correctly aligned 
when 
using \natalie, whereas the other query strategies require at least 800 queries 
for obtaining a perfect alignment. For social networks 
(Figure~\ref{fig:res1}; middle) \margin clearly outperforms \lccl, while \gm and \tm 
outperform both of them except for when querying 26 nodes or more and using 
\tm with \netalign. 
Finally, in family trees (Figure~\ref{fig:res1}; bottom), \gm and \tm consistently 
outperform the 
other methods when using \natalie. With \netalign, \lccl and \margin 
outperform the sampling methods after querying half of the nodes, and in 
contrast to the other experiments, \gm performs clearly worse than \tm within 
the 
first half of the queries. To understand this difference better, we ran \gm 
with 
different temperatures and noticed that with $\beta = 0.001$, \gm obtains a 
comparable performance with \tm. This suggests that the optimal $\beta$ value 
is sometimes problem dependent.

A surprising observation is that in most cases \betw is outperformed even 
by random 
querying. This is probably explained by the fact that \betw queries central 
nodes that have many neighbors and are thus less uncertain since the neighbors 
help inferring their correct alignment.

We conclude that the best overall performance is obtained using the \gm and \tm 
methods; 
\lccl and \margin are both competitive baselines but in some cases they yield 
more than 15 percentage point lower accuracies, whereas the sampling methods 
perform consistently well.

\subsection{Batch querying and scalability}

\begin{figure}[!t]
	\centering
	\includegraphics[width=\columnwidth]{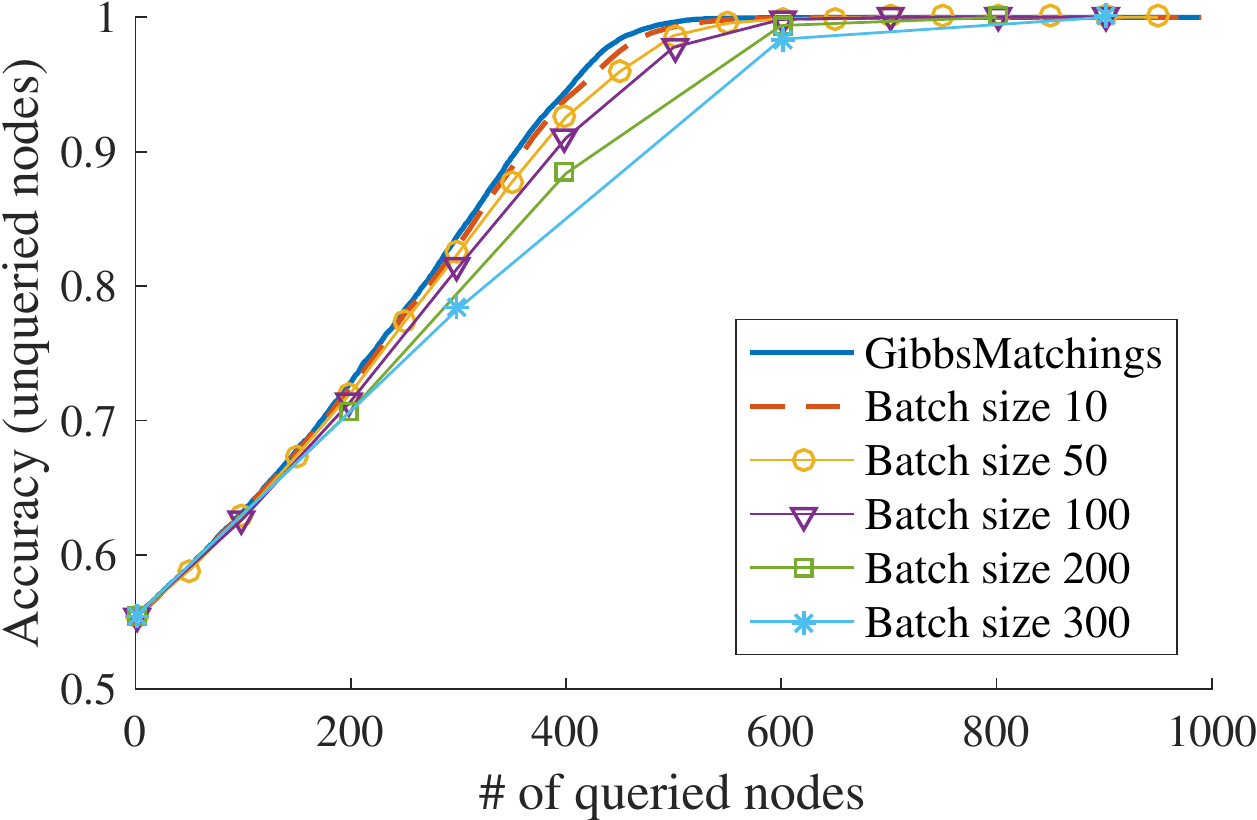}
	\caption{The effect of querying nodes in batches to the overall alignment 
	accuracy.}
	\label{fig:batching}
\end{figure}

When aligning large networks with the help of human experts, getting the 
responses from the humans easily becomes the bottleneck of the algorithm. 
To circumvent this problem, we can query batches of nodes
as discussed in Section~\ref{sec:batch}.
To study the effect of batch querying on \gm, we run an experiment on the 
preferential-attachment graphs and vary the batch size.
Figure~\ref{fig:batching} shows the effect of the batch size on the 
alignment accuracy.
With a batch size of 10, the accuracy is hardly affected, and even when 
batch size is increased to 100 nodes, the accuracy decreases by no more than 
\p{3.5}. This shows that we can obtain accurate
network alignments even when querying the experts in parallel.

Another potential bottleneck is computing the nodes to query. 
Table~\ref{tab:time} lists the running times for computing a query node using 
\tm and \gm combined with \natalie, when aligning two preferential-attachment 
graphs with a varying number of nodes. The number of matchings is set to 
$\ell=30$ with both methods (note that in the main experiments, we used 3\,000 samples with \gm). With networks of up to 1\,000 nodes, \tm is not yet 
a significant bottleneck but when the number of nodes becomes 10\,000, \tm takes already 2.6 
hours.\footnote{We use the original algorithm proposed by Murty \cite{murty68algorithm} for \tm instead of the optimization by Miller et al.~\cite{miller1997}. However, instead of the standard Hungarian algorithm, we employ a bipartite matching solver optimized for sparse graphs, available at: \url{https://www.cs.purdue.edu/homes/dgleich/codes/netalign/}} \gm, on the other hand, scales well so that and even for graphs with 
100\,000 nodes it only takes 10 minutes to sample 30 matchings. Furthermore, 
\gm can be parallelized in a straightforward manner by running multiple 
independent Markov chains simultaneously.

\begin{table}[tb]
	\caption{Running times for the proposed query methods and for the network alignment method \natalie.}\label{tab:time}
	\centering
	\resizebox{\columnwidth}{!}{  
	\begin{tabular}{>{\raggedleft}m{0.17\columnwidth} >{\raggedleft}m{0.25\columnwidth} >{\raggedleft}m{0.28\columnwidth} >{\raggedleft}m{0.15\columnwidth} m{0\columnwidth}}
		\toprule
		Number of \newline nodes & \tm \newline query time & \gm \newline query 
		time & 
		Alignment \newline time & \\ 
		\midrule
		100 & 1.2 sec & 0.03 sec & 3.9 sec & \\
		1\,000 & 21.7 sec & 0.13 sec & 11.0 sec & \\
		10\,000 & 9240.5 sec & 5.10 sec & 48.1 sec & \\
		100\,000 & -- & 621.90 sec & 372.9 sec & \\
		\bottomrule
	\end{tabular}
	}
\end{table}

Finally, we also discovered that \natalie can be optimized when employing it 
multiple times for a sequence of similar problems. More specifically, after we 
have queried a node and fixed its alignment, otherwise keeping the 
problem unaltered, we can leverage the solution of the previous problem for 
solving the current problem. This is achieved by initializing the Lagrangian 
multipliers 
corresponding to the relaxed constraints by the optimal values of the multipliers 
in the previous run of \natalie. In our initial experiments, this strategy 
provided speed ups of more than \p{60} for the convergence of the subgradient 
optimization. However, since \natalie is typically not the bottleneck of the 
proposed active network alignment approach, we have omitted these results.

\section{Conclusions}
\label{section:conclusions}

In this paper
we formalized an active-learning framework, 
which allows us to incorporate human feedback into the 
network alignment problem (also known as graph matching).
In this framework, we obtain human feedback by asking relative queries
to make interaction with experts easier, 
whereas most of the existing works 
on active network alignment rely on absolute queries.
Moreover, we develop a scheme for selecting which nodes to obtain human feedback for. 
Our approach relies on sampling a set of matchings in a transformed bipartite graph, 
and quantifying the certainty of each node using the marginal distribution of the node.
We study two alternative methods for 
sampling matchings, 
\gm, where matchings are sampled according to their score, and 
\tm, where the top-$\ell$ matchings are taken in the sample.
The two proposed methods are shown to 
outperform several previously-proposed baseline methods, 
while offering a robustness-scalability trade-off.

This study opens many possibilities for future work. First, samples of 
matchings could be used for not only relative but also for absolute queries. It 
would be interesting to perform an experimental comparison between the two 
query strategies (absolute vs.\ relative) 
but in order to quantitatively compare their effectiveness, we 
would need to develop a model to compare the cognitive load induced by the 
different type of human expert queries. 
Second, in this work we assume the oracle to always provide the correct alignment for a node but it would be useful to study how the results change given an imperfect oracle.
Third, it would be useful to develop methods for tuning the 
temperature parameter $\beta$ in \gm so that \gm would generalize better to
different types of networks. One possibility would be to compute the maximum 
likelihood estimate for $\beta$ based on the results of the previous oracle 
queries and adaptively update the estimate when new oracle results are 
received.

\section*{Acknowledgments}
We would like to thank Pekka Valta and the Genealogical Society of  
Finland for providing the data for the family-tree experiment, 
as well as the anonymous reviewers for constructive comments and suggestions.
Eric Malmi and Aristides Gionis were supported by the Academy of Finland project ``Nestor''
(286211) and the EC H2020 RIA project ``SoBigData'' (654024).
Evimaria Terzi was supported by a Nokia Visiting Professor Fellowship and NSF grants:
IIS 1320542, IIS 1421759, and CAREER 1253393.

\bibliographystyle{ACM-Reference-Format}
\bibliography{active_network_alignment-short}

\end{document}